\documentclass[
    ,final            
  ]
  {aipproc}

\layoutstyle{6x9}

\newcommand{\circp}{0}
\newcommand{\E}[1]{\mbox{$\times$10$^{#1}$}}
\newcommand{\EA}[1]{\mbox{10$^{#1}$}}

\newcommand{\Tav}{\langle T\rangle}

\newcommand{\et}{\mbox{$\eta$}}
\newcommand{\etp}{\mbox{$\eta'$}}
\newcommand{\po}{\mbox{$\pi^0$}}
\newcommand{\pio}{\mbox{$\pi^0$}}
\newcommand{\gm}{\mbox{$\gamma$}}
\newcommand{\pim}{\mbox{$\pi^-$}}
\newcommand{\pip}{\mbox{$\pi^+$}}
\newcommand{\elp}{\mbox{$e^+$}}
\newcommand{\elm}{\mbox{$e^-$}}

\begin{document}

\title[{\bf DRAFT v1.7}]{What is interesting in $\eta$ and $\eta'$ Meson 
Decays?
}

\author{  Andrzej  Kup\'s\'c}{   address={
Institutionen f\"or K\"arn- och Partikelfysik, Uppsala University, Uppsala,
SWEDEN
\\
The Andrzej So{\l}tan Institute for Nuclear Studies, Warsaw, POLAND.
}
}
\begin{abstract}
An  introduction to  the physics  of \et\  and \etp\  meson  decays is
given.   A  historical account  of  the  discovery  of the  mesons  is
presented.  It  is followed by  an overview and classification  of the
common decay modes  and the relevance of the  mesons for modern hadron
and particle  physics.  In  more detail the  hadronic decay  modes are
discussed and in particular  some interesting features of the $\eta\to
3\pio$  decay are  presented.  The  last section  briefly  reviews and
compares reactions used  to produce the \et\ and  \etp\ mesons for the
studies of their decays.
\end{abstract}
\maketitle
\section{Historical note}

The  $\eta$   meson  was  discovered   in  1961  by  Pevsner   et  al.
\cite{Pevsner:1961pa} (paper submitted 10th November) in the year when
{\em  The Eightfold  Way} was  formulated and  the $\omega$  meson was
discovered.   The $\eta'$  meson was  discovered independently  by two
groups in  1964: Kalbfleisch et  al.  \cite{PhysRevLett.12.527} (paper
submitted  9th April) and  Goldberg et  al.  \cite{PhysRevLett.12.546}
(paper  submitted  15th April).  In  that  year  the quark  model  was
proposed and the $\Omega^-$ $S=3$ particle was discovered.

The  earliest predictions of  $\eta$ and  $\eta^\prime$ can  be traced
back to  works within the Sakata  model~\cite{Sakata:1956hs} where the
nucleons and the  $\Lambda$ particle were used as  the building blocks
(so called sakatons). In this scheme the isospin 0 neutral mesons were
needed to accompany  pions and kaons, the only  pseudoscalars known at
that  time. The  new  mesons  were decomposed  as  follows (Okun  1957
\cite{Okun:1957aa,       Okun:1958aa},        Ikeda       et       al.
\cite{1959PThPh..22..715I}  and  Yamaguchi  \cite{1958PThPh..19..622Y,
1959PThPS..11....1Y})\footnote{In  modern terminology: $\pi^{0\prime}$
-- \et; $\pi^{0\prime\prime}$ -- \etp.}: $\pi^{0\prime\prime}= \Lambda
\bar\Lambda \;  , \;\;  \pi^{0\prime}= (p\bar p  - n\bar  n)/\sqrt 2$.
The  relation  between  $\pi^{0\prime}$  and  $\pi^{\pm,0}$  would  be
entirely   analogous   to   the   relation   between   $\Lambda$   and
$\Sigma^{\pm,0}$.  Since the mesons were an unavoidable consequence of
the  approach   their  absence  was   used  to  criticize   the  model
(Sakurai~\cite{Sakurai:1960ju} 1960).

In {\em The  Eightfold Way} Gell-Mann refers to  \et\ (named $\chi^0$)
in the following phrase (\cite{Gell-Mann:1961ky}, March 15, 1961):
\begin{quote}
... The most  clear-cut new prediction for the  pseudoscalar mesons is
the  existence of  $\chi^0$, which  should decay  into  2$\gamma$ like
$\pi^0$, unless it is  heavy enough to yield $\pi^++\pi^-+\gamma$ with
appreciable   probability.  (In   the  latter   case,  we   must  have
($\pi^+\pi^-$)  in an  odd state.)  $\chi^0\to 3\pi$  is  forbidden by
conservation of I  and C. For a sufficiently  heavy $\chi^0$ the decay
$\chi^0\to 4\pi$ is possible, but hampered by centrifugal barriers.
\end{quote}

The events  which lead  to the \et\  discovery and  interpretation are
nicely summarized by Rosenfeld in an account of the early years of the
Particle Data Group (PDG) \cite{Rosenfeld:1975fy}:
\begin{quote}
 A meson  (apparently unrelated because  it decayed into  three pions)
\cite{Pevsner:1961pa}  was  discovered  late  in  1961,  and  properly
identified  \cite{PhysRevLett.8.114}  as  the  predicted  pseudoscalar
meson,  the $\eta$,  early in  1962.  This completed  the first  meson
octet, but by latter standards it attracted little attention (no press
conference, no  flurry of theoretical  papers, and no 1962  edition of
UCRL-8030\footnote{Precursor of  the PDG reports first  issued in 1957
\cite{Barkas:1957aa} and than often updated.}).
\end{quote}
In  a  retrospective   on  the  origins  of  the   quark  model  Zweig
\cite{Zweig:1980nu} comments on the fate of the predictions within the
Sakata model: `` ...Their striking  prediction of the existence of the
$\eta$ was  not mentioned by  the experimental groups  that discovered
and studied  this key particle''.  Finally Okun  summarizes the events
in the year 1964 in the following way \cite{Okun:2006nq}:
\begin{quote}
 In  1964  the   $\eta^\prime$-meson  and  the  $\Omega$-hyperon  were
discovered \cite{PhysRevLett.13.449,Barnes:1964pd}.  Earlier this year
G.~Zweig \cite{Zweig:1964wu} and M.  Gell-Mann \cite{Gell-Mann:1964nj}
replaced  the   integer  charged  sakatons   by  fractionally  charged
particles (aces -- Zweig; quarks  -- Gell-Mann).  This allowed them to
construct not only the octet and singlet of mesons, but also the octet
and decuplet of baryons.
\end{quote}

\section{Main decay modes}

\begin{figure}
  \hfill\includegraphics[height=.15\textheight]{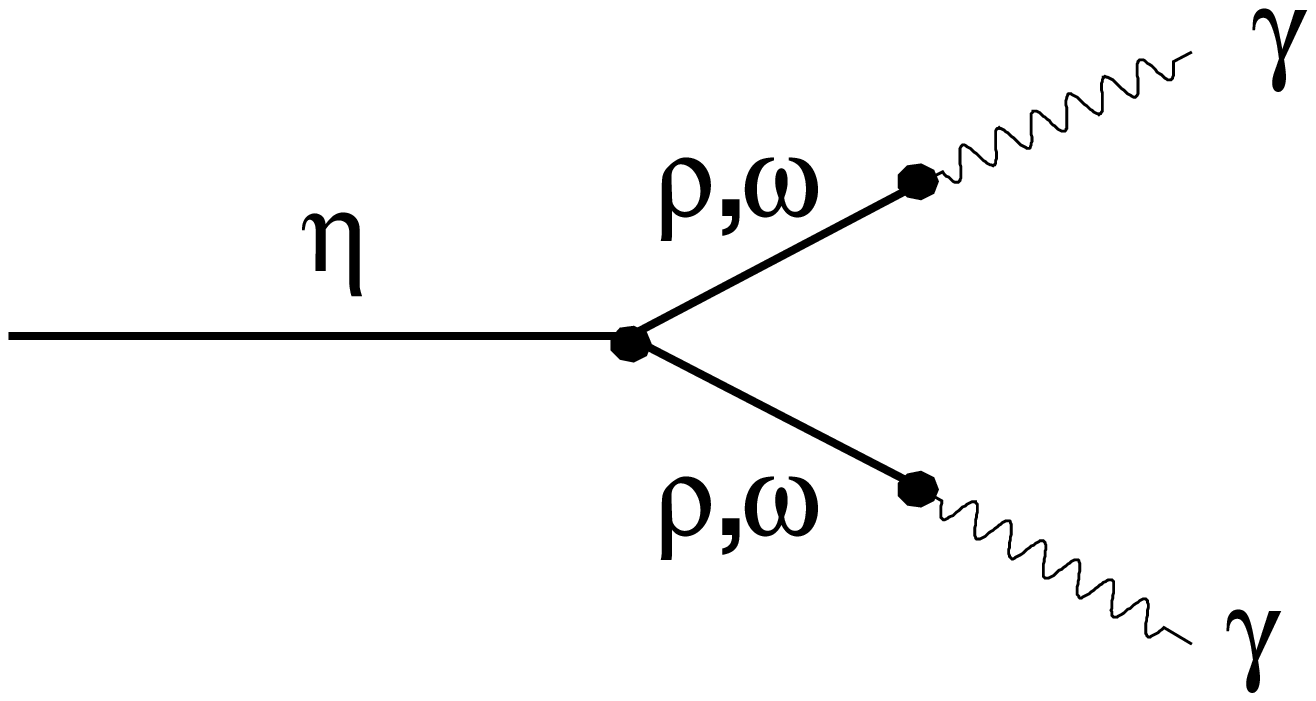}
  \includegraphics[height=.15\textheight]{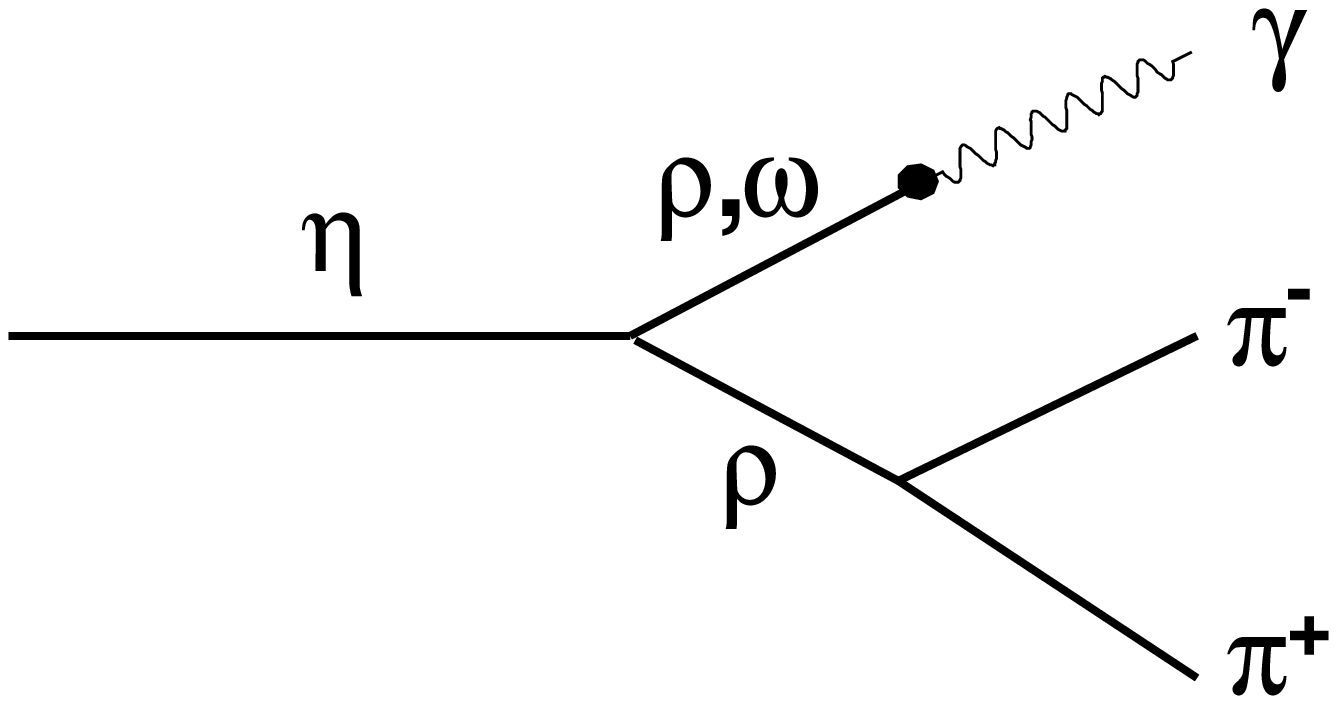}\hfill
  \caption{\label{fig:VMDgraph}Vector  Meson dominance  model  for eta
  anomalous decays.}
\end{figure}

The basic facts  about the \et\ and \etp\ mesons  from the most recent
issue   of  the   PDG  report~\cite{Yao:2006px}   are   summarized  in
table~\ref{tab:eprop}.   The  main decays  fall  into two  distinctive
classes: hadronic decays into  three pseudoscalar mesons and radiative
decays.   This fact  was  pointed  out already  in  1962 by  Gell-Mann
\cite{Gell-Mann:1962jt} in case of the \et\ meson decays:
\begin{quote} 
The  forbidden  decay  rates  into 3$\pi^0$  and  $\pip+\pim+\po$  are
difficult   to   estimate,   except  that   $3\pi^0/(\pip+\pim+\po)\le
\frac{3}{2}$.  The remaining neutral  decays are expected, however, to
represent  $\chi\to  2\gamma$  The  decay  $\chi\to  2\gamma$  may  be
described roughly  on the  assumption that the  important intermediate
steps   are   $\chi\to   2\rho^0$  (followed   by   $\rho^0\to\gamma$,
$\rho^0\to\gamma$)     and    $\chi\to    2\omega$     (followed    by
$\omega\to\gamma$,  $\omega\to\gamma$).  We now  wish to  estimate the
ratio of this  rate to that of hitherto  unobserved charged decay mode
$\chi\to\pip+\pim+\gamma$,  which  should  be  dominated  by  $\chi\to
2\rho^0$ followed by $\rho^0\to\gamma$, $\rho^0\to\pip+\pim$.
\end{quote}
The description  proposed here  for the radiative  decays is  based on
Vector  Meson Dominance  (VMD)  model formulated  by  Sakurai in  1960
\cite{Sakurai:1960ju}     and    represented  by  the   diagrams    in
Fig.~\ref{fig:VMDgraph}. Using the  intermediate state with two vector
mesons one  can describe at  least qualitatively the main  features of
the  radiative decays.   This approach  fails however  when  trying to
describe  decays   into  three  pseudoscalars.   The  $\eta,\eta'  \to
\pi\pi\pi$ decays do not  conserve isospin since Bose symmetry forbids
the three pions with $J^P=0^-$ to occur in the isoscalar state. If one
acknowledge this fact and consider  that the final state has isospin 1
then  a  possible  intermediate  state  could  be  $\pi^\pm  \rho^\mp$
($\pi^0\rho^0$ is  forbidden by $C$ conservation).   The $\pi^\pm$ and
$\rho^\mp$ have  to be in a relative $p$-wave and cannot contribute to
the neutral decay mode into  three \pio. This is obviously in contrast
with the experiment since the neutral mode is 1.41 times more frequent
than  $\pip\pim\pio$  and  close  to  the  upper  boundary  quoted  by
Gell-Mann.   The  solution  would  be  to assume  that  the  two  body
intermediate states include also scalar mesons. The scalar mesons were
discovered  later  and  even  today  their properties  are  not  well
established.  Therefore the  hadronic decays  of \et\  and  \etp\ into
three pseudoscalar mesons  provide a source of information  on the low
energy scalar interactions.

\begin{table}
\begin{tabular}{ll|ll}
\hline
$M_\eta=547.51\pm0.18$ MeV&&$M_{\eta'}=957.78\pm 0.14$ MeV\\
$\Gamma_\eta=1.30\pm 0.07$ keV && $\Gamma_{\eta'}=0.203\pm 0.016$ MeV\\
$\eta\to\gamma\gamma$    &39\%&$\eta'\to \pi^+\pi^-\eta$&44\%\\
$\eta\to\pi^0\pi^0\pi^0$ &32\%&$\eta'\to \rho^0\gamma$  &29\%\\
$\eta\to\pi^+\pi^-\pi^0$ &23\%&$\eta'\to \pi^0\pi^0\eta$&21\%\\ 
$\eta\to\pi^+\pi^-\gamma$& 5\%&$\eta'\to \omega\gamma  $& 3\%\\
&&  $\eta'\to \gamma\gamma$               & 2\%\\
\hline
\end{tabular}
\caption{\label{tab:eprop}Main properties of the \et\ and \etp\ mesons
\cite{Yao:2006px}.}
\end{table}

\section{Modern perspective}

The \et\  and \etp\ mesons plays  a special role  in understanding low
energy Quantum  Chromodynamics (QCD).   They are isoscalar  members of
the nonet of the  lightest pseudo-scalar mesons.  The systematical way
of dealing with that regime  of QCD is provided by Chiral Perturbation
Theory (CHPT)  \cite{Gasser:1983yg}. The global symmetry  of QCD where
the  three  light  quarks  are  massless is  the  chiral  symmetry  --
$SU(3)_L\times  SU(3)_R$ which is  spontaneously broken  to $SU_V(3)$.
The octet  of Goldstone pseudoscalar  mesons, arising from  the broken
symmetry, is identified with pions,  kaons and eta.  The particles are
used as  degrees of  freedom in CHPT  and a systematical  expansion is
performed in  powers of momenta of  the mesons.  At each  level of the
calculations one has to introduce new counterterms (free parameters of
the theory) in order  to restore renormalizability. Direct breaking of
the chiral  symmetry is  accounted by expanding  in the  quark masses.
Standard  CHPT provides  an  accurate description  of  the strong  and
electroweak interactions  of the  pseudoscalar mesons at  low energies
\cite{Bijnens:2004pk, Ecker:2005za}.   Due to weak  $K_L-\et$ mixing a
good understanding of $\eta$  decays is a prerequisite for calculation
of  Standard   Model  contributions  to  rare  kaon   decays  such  as
$K_L\to\po\elp\elm$ \cite{Ecker:1987qi,Flynn:1988gy}.  CP violation in
the flavor conserving processes can be tested in the \et\ decays which
are analogous  to those of  $K_L$: $\et\to\pi\pi$, $\et\to\po\elp\elm$
and in the angular asymmetry between the $\pip\pim$ and the $\elp\elm$
decay planes in the $\et\to\pip\pim\elp\elm$ decay.  Conversion decays
of \et\  provide information about the  pseudoscalar formfactor needed
for example for light-by-light  contribution to the anomalous magnetic
moment of the  muon.  It was also pointed  out that $\et,\et'\to 3\pi$
decays provide valuable information on light quark masses.

Most of the complications of the $\eta$ treatment in the standard CHPT
are due to mixing of the $\eta$ and $\eta'$ fields.  The \etp\ is most
esoteric meson of the pseudoscalar  nonet, since it is closely related
to the  axial $U(1)$  anomaly and remains  massive even in  the chiral
limit.   It was observed  by Witten  \cite{Witten:1979vv} that  in QCD
with  infinite  number of  colors  \cite{'tHooft:1973jz} the  $\eta_0$
state is  massless and the global $SU(3)_L\times  SU(3)_R$ symmetry is
replaced by $U(3)_L\times U(3)_R$. Corresponding effective Lagrangians
were   studied  by  e.g.    \cite{Veneziano:1979ec,  DiVecchia:1980ve,
Rosenzweig:1979ay, Kawarabayashi:1980dp}.  A systematical treatment of
the $1/N_c$ expansion was introduced into CHPT by Kaiser and Leutwyler
\cite{Kaiser:2000gs}.   An  alternative  approach  is $U(3)$  CHPT  of
Beisert and  Borasoy \cite{Beisert:2001qb}.  With that  tool mixing of
$\eta$  and $\eta'$ has  received a  solid theoretical  foundation. In
particular it  has been recognized both from  phenomenology and theory
side that two mixing angles are required to describe the mixing in the
singlet   and  octet  basis   \cite{Schechter:1992iz,  Kiselev:1992ms,
Feldmann:1999uf, Escribano:2005qq, Benayoun:2003we}.

Due to the large mass of  $\eta'$ light vector and scalar mesons could
be produced  in the  decays.  Importance of  vector mesons is  seen in
radiative       decay       modes;      $\eta'\to\rho\gamma$       and
$\eta'\to\omega\gamma$.    Contributions   of   light  scalar   mesons
$\sigma,f_0,a_0$  should play a  significant role  in the  decays into
$\eta\pi\pi$ and $\pi\pi\pi$ \cite{Fariborz:1999gr,Abdel-Rehim:2002an}
but it  is not  as apparent  due to the  width of  the mesons  and the
offshell  behavior.  Description  of vector  and scalar  resonances is
beyond the scope of the standard CHPT but there is continuous progress
in  the theoretical  treatment.  For  example the  VMD model  is being
incorporated into the  theory \cite{Fujiwara:1984mp, Ecker:1988te} and
unitarity   constrains  are   implemented   by  dispersion   relations
\cite{Kambor:1995yc,Anisovich:1996tx}.

In  conclusion in  the  early  stages of  CHPT  development, \et\  and
specially  \etp\ were  treated  as an  additional  complication or  by
hidding   their    effects   were    hidden   in   the    low   energy
constans~\cite{Gasser:1984gg}.  Refinements and extensions of the CHPT
methods stimulated by the need  for precise calculations for kaons and
eta  decays provided tools  which enable  to include  the \etp\  in an
elegant and  consistent way and  to perform reliable  calculations for
the processes.  One should  mention a recent theoretical programme for
simultaneous  treatment of  \et\ and  \etp\ decays  within  the chiral
unitary approach  \cite{Beisert:2001qb, Borasoy:2004qj}.  It  is based
on  chiral  perturbation  theory   and  the  unitarization  using  the
Bethe-Salpeter equation.   It is consistent with  the constraints from
chiral  symmetry and the  chiral anomaly,  while at  the same  time it
matches to the one loop expansion of CHPT.  It is discussed by Borasoy
and Ni{\ss}ler in the contributions to this Symposium.

In addition  to the  contributions to the  present Symposium  a recent
introduction  to the  topic of  \et\ decay  physics is  also  given by
Bijnens~\cite{Bijnens:2005sj} and  the extensive reviews  of the field
could    be   found    in   the    proceedings   of    two   workshops
\cite{Bijnens:2002zy, ETA:2005aa}.

\section{Hadronic decays}

The three body decays into pseudoscalar mesons constitute the majority
of all  $\eta$ decays -- 55\% ($\eta\to\pi\pi\pi$)  and $\eta'$ decays
-- 65\% ($\eta'\to\pi\pi\eta$) as  seen in table~\ref{tab:eprop}.  The
decays  $\eta,\eta' \to  \pi\pi\pi$ do  not conserve  isospin  and the
partial  widths  are  similar  to  the  second  order  electromagnetic
processes (decays  into $\gamma$ pairs).  It appears however  that the
contributions   from    electromagnetic   processes   are   suppressed
\cite{Sutherland:1967vf, Bell:1968mi, Baur:1995gc}  and the decays are
driven by an isospin violating term of the QCD Lagrangian proportional
to $m_d - m_u$.
\begin{figure}
\includegraphics[width=0.32\textwidth,clip]{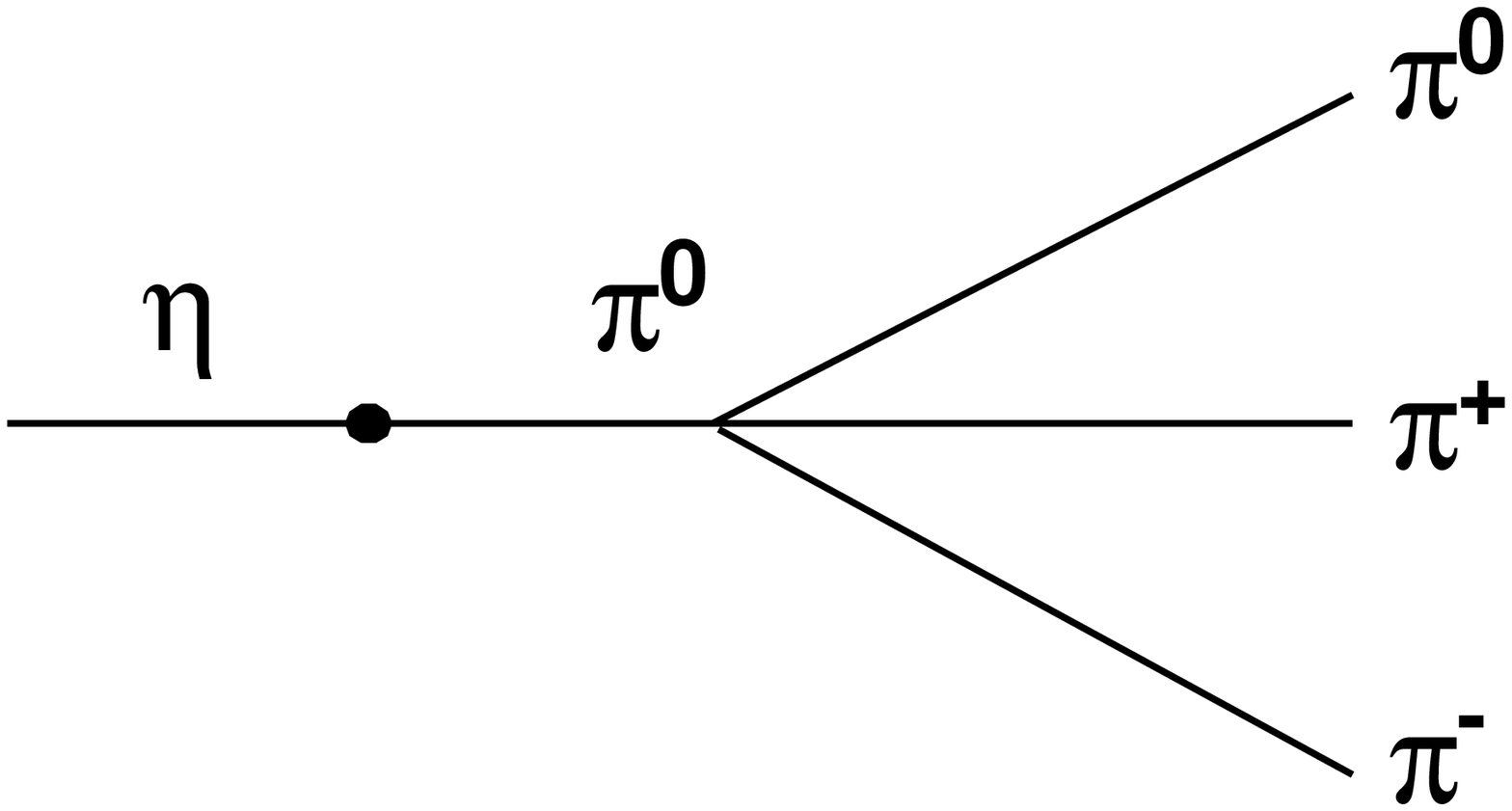}
\includegraphics[width=0.32\textwidth,clip]{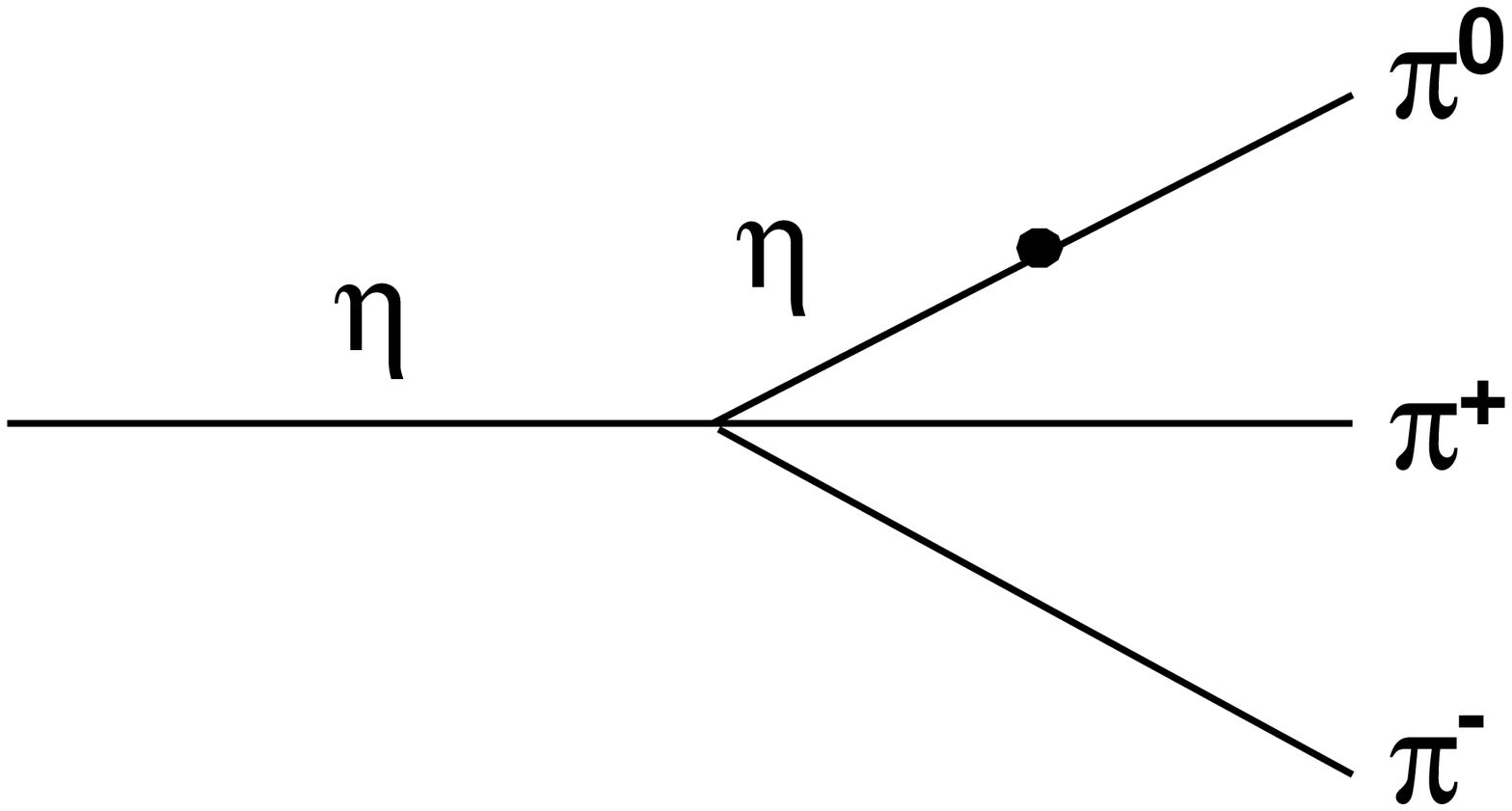}
\includegraphics[width=0.32\textwidth,clip]{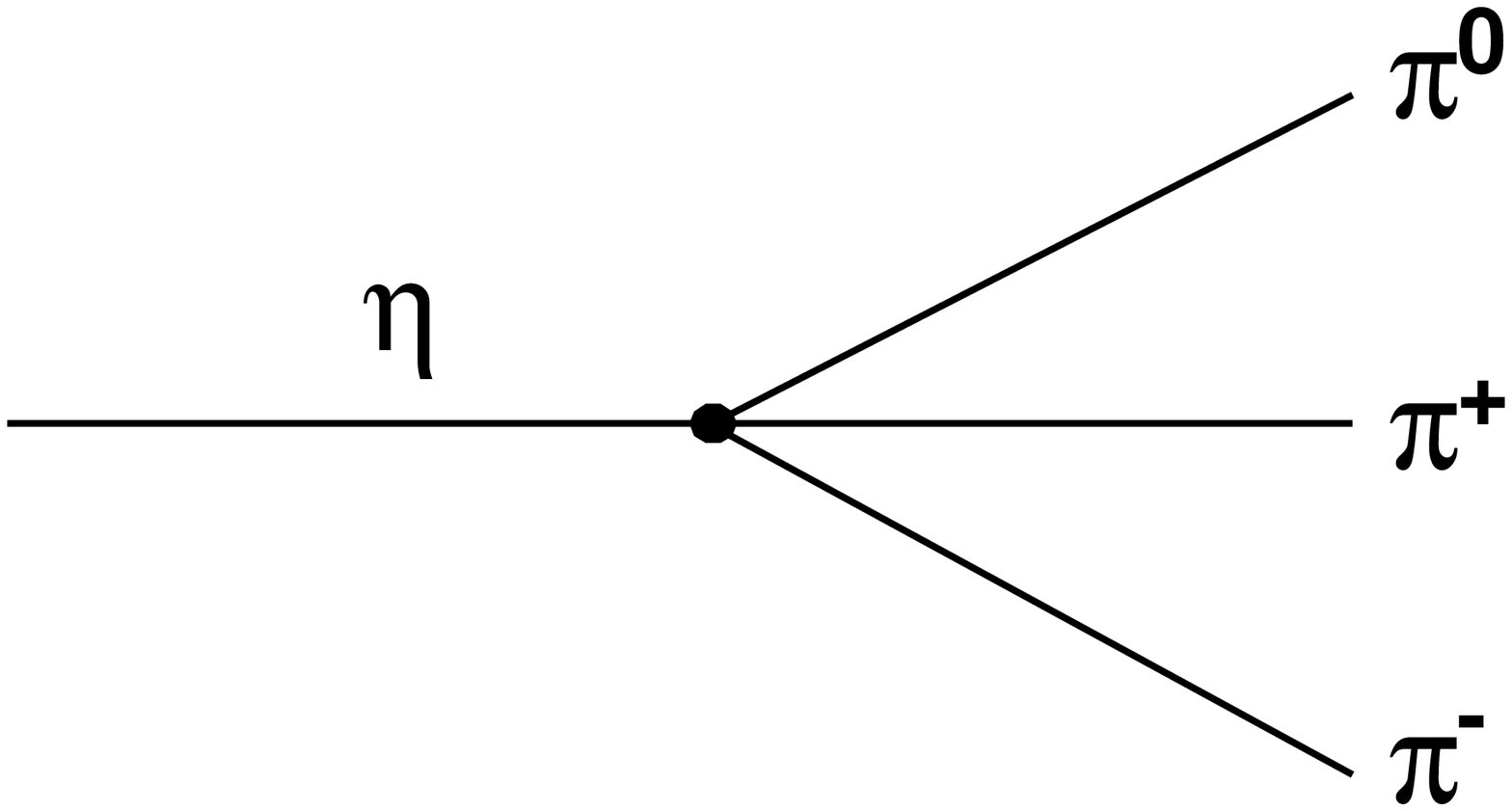}
\caption{\label{fig:grapheta3pi} Isospin violation in $\eta\to
3\pi$. Lowest order effective lagrangian contribution.}
\end{figure}
The lowest order  contribution to the decay mechanism  is given by the
Current    Algebra   graphs   shown    in   Fig.~\ref{fig:grapheta3pi}
\cite{Osborn:1970nn} consisting  of a combination  of \et-\pio\ mixing
and  elementary   low  energy  QCD   process  --  scattering   of  two
pseudoscalar mesons.   The partial width of  the $\eta\to \pi^{\circp}
\pi^+\pi^-$  decay   calculated  using   current  algebra  is   66  eV
\cite{PhysRevLett.18.1170},   much   below   the  experimental   value
294$\pm$16 eV  \cite{Yao:2006px}.  The second order in  the low energy
expansion of the effective Lagrangian  of QCD was calculated by Gasser
and   Leutwyler   in   1984   increasing   the  result   to   160   eV
\cite{Gasser:1984pr}.  The  big change  implies the importance  of the
$\pi\pi$ interaction in  the final state.  One may  expect that higher
loop calculations  enhance the prediction for the  decay further since
they  should give  a better  description of  the final  state $\pi\pi$
interactions.  However,  such calculations would be  very involved and
the result might  not be predictive since it  requires many low energy
constants which  are not well  known.  An elegant method  of including
the  $\pi\pi$  interactions  up   to  higher  orders  is  provided  by
dispersion  relations which connect  the imaginary  part of  the decay
amplitude  with the amplitude  itself.  The  amplitude as  an analytic
function  is uniquely characterized  by its  singularities and  can be
represented  by an integral  over its  discontinuities along  a branch
cut.  The unitarity and  analyticity will determine the amplitude only
up  to some subtraction  polynomial which  could be  constrained using
CHPT.    There    are   two   calculations    using   this   technique
\cite{Anisovich:1996tx,   Kambor:1995yc}   but   employing   different
formalism.  They lead consistently to an enhancement of the decay rate
by about 14\%.  The dependence of the decay width ($\Gamma$) on $m_d -
m_u$ can be expressed in the following way:
\begin{equation}
\Gamma\propto\left(\frac{1}{Q}\right)^4
\end{equation}
where 
\begin{equation}
\frac{1}{Q^2} = \frac{m_d^2-m_u^2}{m_s^2 - \frac{1}{4}(m_d + m_u)^2}.
\label{decay3p}
\end{equation}
The  standard  way  to determine  $Q$  is  to  use the  leading  order
expressions for  the masses of the pseudoscalar  mesons, together with
Dashen  theorem  \cite{Dashen:1969eg}, which  leads  to the  following
formula:
\begin{equation}
Q^2\approx Q_D^2\equiv\frac{m_K^2}{m_\pi^2}
\frac{m_K^2-m_\pi^2}{m_{K^0}^2-m_{K^+}^2+m_{\pi^+}^2-m_{\pi^\circp}^2}
\end{equation}
-- numerically $Q_D$=24.1.  However $\Gamma$ is sensitive to the exact
value of $Q$ (an uncertainty of 10\% in the decay rate translates to a
2.5\% error in  $Q$) and the decay might  provide a precise constraint
for the  light quark mass ratios  \cite{Leutwyler:1996qg}.  Namely $Q$
determines the major  axis of the ellipse in  the $m_u/m_s$, $m_d/m_s$
plane. One rewrites Eqn.~\ref{decay3p} as:
\begin{equation}
\Gamma=\left(\frac{Q_D}{Q}\right)^4\bar{\Gamma}
\end{equation}
where  $\bar{\Gamma}$  is  the  decay  width calculated  in  the  CHPT
assuming  $Q_D=Q$.  Deviation  of $Q$  from  $Q_D$ can  be studied  by
comparing $\bar{\Gamma}$  to the value of the  $\Gamma$ extracted from
experiment.   One important  prerequisite  is the  reliability of  the
$\bar{\Gamma}$  determination.   The  calculations  can be  tested  by
comparing  the ratio  $\Gamma(\eta \to  3\pi^{\circp})/\Gamma(\eta \to
\pi^+\pi^-\pi^{\circp})$  and  the  kinematical distributions  of  the
pions with  experiment.  Recently $Q$  was derived by  determining the
subtraction  polynomial  within the  dispersion  relation approach  of
\cite{Kambor:1995yc}    from   preliminary    KLOE    data   on    the
$\eta\to\pi^+\pi^-\pi^0$        decay        \cite{Giovannella:2005rz,
Ambrosino:2007mq},      yielding      the      $Q$     value      22.1
\cite{Martemyanov:2005bt}.  In an  alternative approach by Borasoy and
Ni{\ss}ler   \cite{Borasoy:2005du}  rescattering   of   any  pair   of
pseudoscalar mesons is taken  care of by the Bethe-Salpeter equations.
The parameters beyond the one  loop are constrained by fitting both to
pseudoscalar  scattering data and  to the  \et, \etp\  reactions.  The
conclusion is  that the current experimental situation  does not allow
to  constrain the  parameters, in  particular to  conclude  that $Q\ne
Q_D$, and  therefore to provide precise information  about light quark
masses within this approach.

\begin{figure}
  \includegraphics[height=.3\textheight]{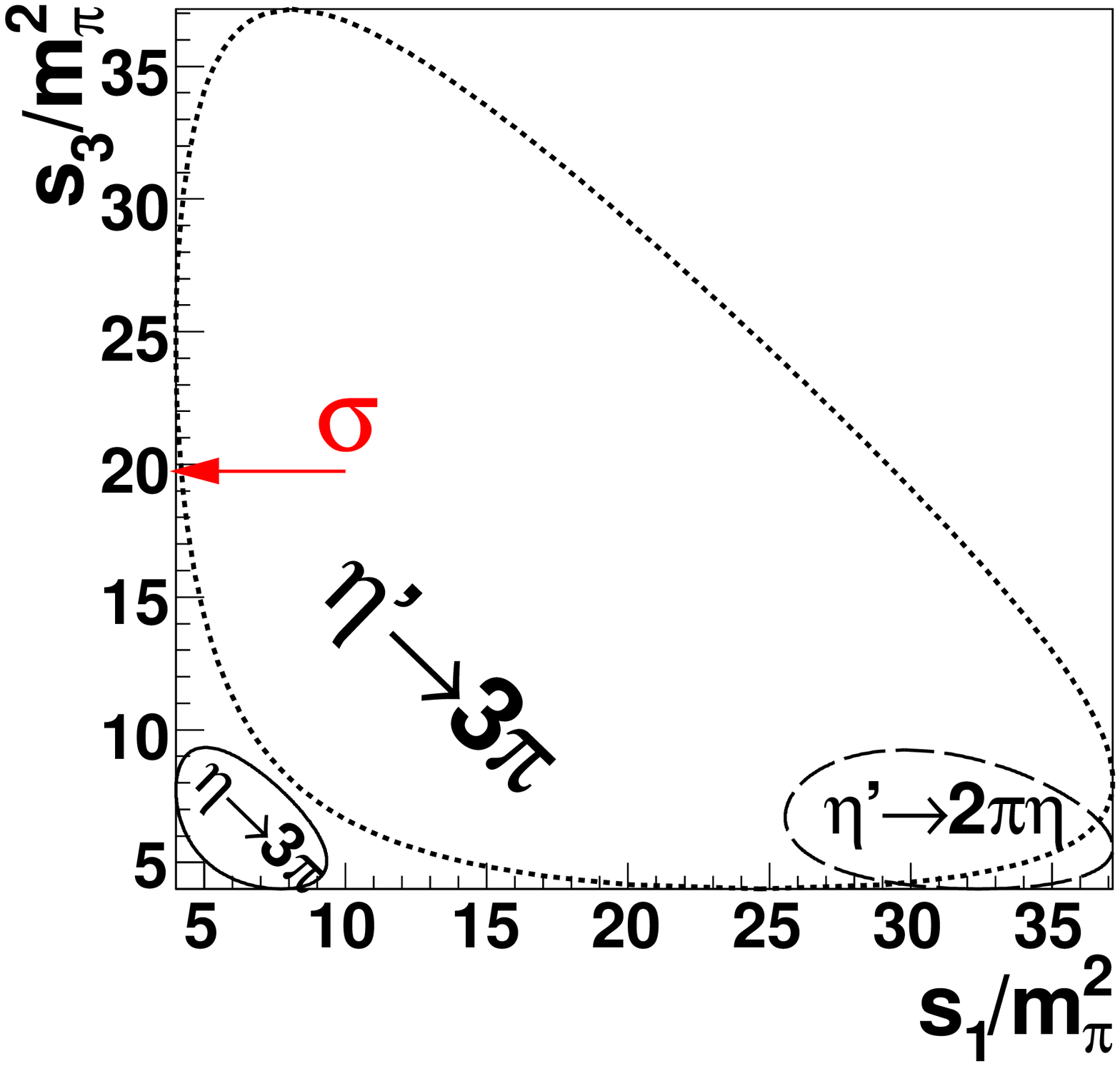}
  \includegraphics[height=.3\textheight]{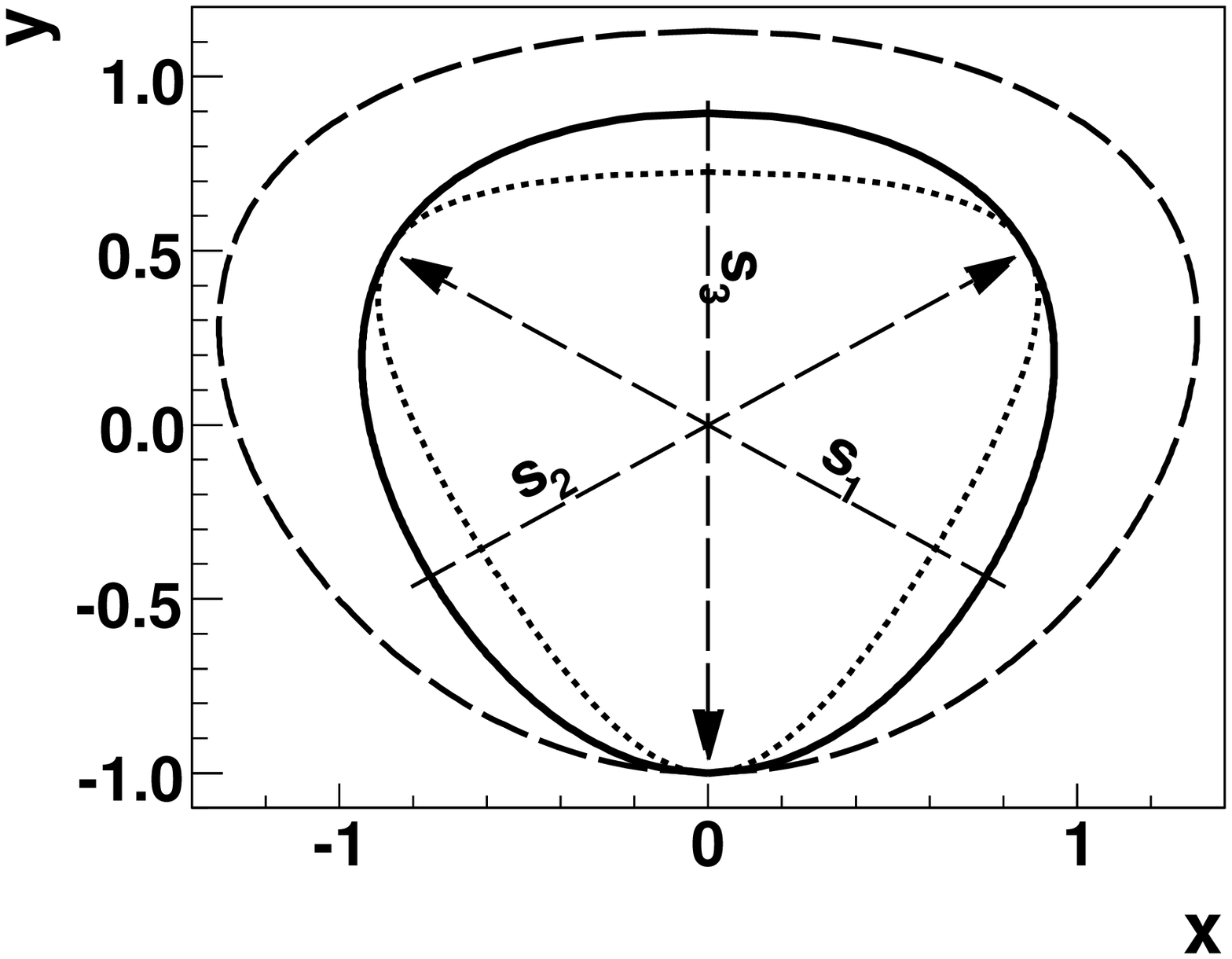}
  \caption{\label{fig:sigdal} Dalitz plot  boundaries for the \et\ and
\etp\   hadronic    decays:   $\eta\to\pi\pi\pi$   --    solid   line,
$\eta'\to\pi\pi\pi$ -- dotted  line and $\eta'\to\pi\pi\eta$ -- dashed
line. (Left) The variables  $s_1$, $s_2$ normalized to $m_{\pi^0}$ are
used.   The  arrow indicates  the  mass  of  the $\sigma$  meson  (600
MeV). (Right) The Dalitz plots in the normalized $x,y$ variables.}
\end{figure}

For  a three  body decay  there are  only two  independent kinematical
variables to describe  the phase space. A convenient  common choice is
to use Mandelstam variables which for the decay $0\to 1+2+3$ ($p_i$ is
the fourmomentum) can be written:
\begin{equation}
s_i=(p_0-p_i)^2.
\end{equation}
This means  that for  example $s_3$ is  the invariant mass  squared of
particles 1 and 2.  The $s_i$  can be expressed by kinetic energies of
the decay products ($T_i$) in the  center of mass system of the mother
particle:
\begin{equation}
s_i=(m_0-m_i)^2 - 2T_im_0.
\end{equation}
The variables are therefore not independent:
\begin{equation}
\sum_{i=1}^3T_i=
m_0-\sum_{i=1}^3m_i\equiv 3\Tav,
\end{equation}
where the introduced variable $\Tav$  is the average kinetic energy of
the  outgoing  particles.   When  any  two $s_i$  are  chosen  as  the
independent  variables the  phase space  can be  represented by  a two
dimensional   plot   (the   Dalitz   plot~\cite{Dalitz:1953cp}).   The
probability density in such variables is proportional to the amplitude
squared for the  process.  The boundaries of the  Dalitz plots for the
\et\ and \etp\ hadronic decays are shown in Fig.~\ref{fig:sigdal}.

In the case when $m_1=m_2\equiv m$ one uses symmetrized, dimensionless
variables to parametrize the phase space:
\begin{equation}
  x\equiv\frac{1}{\sqrt{3}}\frac{T_1-T_2}{\Tav};\ \   
y\equiv\frac{1}{3}\left(\sum_{i=1}^3 \frac{m_i}{m}\right)\frac{T_3}{\Tav}-1.
\end{equation}
The  kinematical   boundaries  in  the   $x,y$  plane  are   shown  in
Fig.~\ref{fig:sigdal}   for   \et\    and   \etp\   hadronic   decays.
Phenomenologically the  mechanism of the hadronic three  body \et\ and
\etp\ decays are  described by an expansion of  the amplitude $A(x,y)$
around  the center  $x=y=0$ point  and  therefore the  density of  the
Dalitz plot can be written as:
\begin{equation}
|A(x,y)|^2\propto 1+ay+by^2+cx+dx^2...
\label{eqn:dalexpans}
\end{equation}
The term linear in $x$ vanishes in  case when 1 and 2 are identical or
are  particle-antiparticle pair  and Charge  Conjugation ($C$)  is not
violated.

One  can see  \cite{Gasser:1984pr} that  the pions  must emerge  in an
$I=1$ configuration  and the following relation  between amplitudes of
the  $A\equiv   A_{\pip\pim\pio}$  and  $\bar{A}\equiv  A_{\po\po\po}$
follows:
\begin{equation}
\bar{A}(s_1,s_2,s_3)=A(s_1,s_2,s_3)+A(s_2,s_3,s_1)+A(s_3,s_1,s_2).
\label{eq:threepiamp}
\end{equation}
The amplitude $A(s_1,s_2,s_3)$ derived  within current algebra (lowest 
order of CHPT) from graphs in Fig.~\ref{fig:grapheta3pi} leads to 
\cite{Osborn:1970nn}:
\begin{equation}
A(s_1,s_2,s_3)=A(s_3)\propto \frac{3s_3-4m_\pi^2}
{m_\eta^2-m_\pi^2}.
\label{eq:pcac}
\end{equation}
The amplitude is a linear function  in $s_3$ (or in $y$).  In terms of
the expansion from  Eqn.~\ref{eqn:dalexpans} the result corresponds to
$a=-1.052$    and    $b=0.28$.    For    the    neutral   decay
Eq.~\ref{eq:threepiamp}  leads  to  the constant  amplitude  $\bar{A}$
always when $A$ is a linear  function of $s_3$.  The result is dominated
by  the  left  graph  in  the  Fig.~\ref{fig:grapheta3pi}  (pion  pole
diagram).

In case of the isospin  conserving decays $\eta' \to \eta\pip\pim$ and
$\eta' \to \eta\pio\pio$ the wave function must be symmetric under the
exchange  of the  pions and  the  three-particle final  state must  be
isoscalar.  This leads to a simple relation:
\begin{equation}
A_{\pip\pim\et}(s_1,s_2,s_3) = A_{\pio\pio\et}(s_1,s_2,s_3) \ .
\end{equation}
This leads to $r_1\equiv\Gamma(\eta' \to \eta \pi^+\pi^-)/\Gamma(\eta'
\to \eta \pi^0  \pi^0)=2$ due to a factor $1/2$ for  the case with two
identical particles.  The mass  difference between charged and neutral
pions leads  to smaller phase  space volume for the  $\pi^+\pi^-$ case
than  for  the $\pi^0\pi^0$  and  decreases  the  ratio to  1.78  $\pm
0.02$~\cite{Borasoy:2005du}.   The ratio $r_1$  has not  been measured
directly  and the evaluation  from the  known branching  rations gives
$2.14 \pm 0.19$~\cite{Yao:2006px}.  Such  a large value would indicate
significant isospin-violating contributions in the amplitude.

As it was  stated previously for a consistent  description of \et\ and
\etp\ decays the final  state interactions and \et-\etp\ mixing should
be  taken into  account.   Such a  complete  and consistent  framework
\cite{Beisert:2002ad,  Borasoy:2005du}  is  presented  also  in  other
contributions to this symposium.

\section{Peculiarities of the  $\eta\to 3\po$ decay}

For decays  with three identical particles $\eta,\eta'\to  3\po$ it is
useful to introduce polar coordinates $(\sqrt{z},\phi)$ in the $(x,y)$
plane to parametrize the phase space:
\begin{equation}
x=\sqrt{z}\sin\phi; y=\sqrt{z}\cos\phi.
\end{equation}
The Dalitz plot  has sextant symmetry and therefore  one can limit the
range of $\phi$ to $0\le\phi<60^\circ$. The variable $z$ is than given
by:
\begin{equation}
z= x^2+y^2=
\frac{2}{3}\sum_{i=1}^3\left(\frac{T_i-\Tav}{\Tav}\right)^2.
\label{eqn:z}
\end{equation}
and $0\le z\le 1$.
\begin{figure}
  \includegraphics[height=.3\textheight]{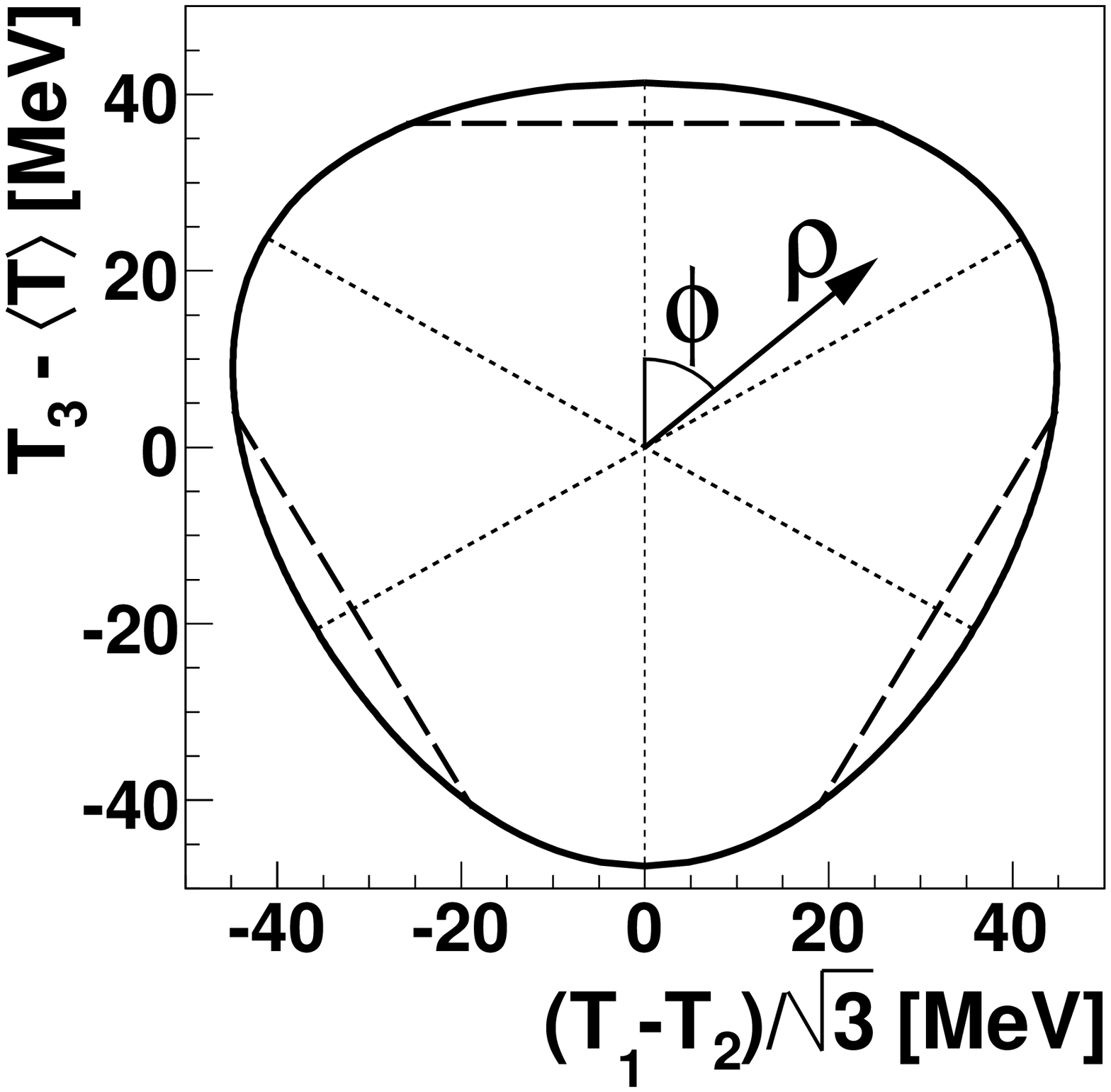}
  \caption{\label{fig:def}Symmetrized Dalitz plot for 
$\eta\to 3\pi^{\circp}$ decay ($\rho\equiv\sqrt{z}\Tav$).}
\end{figure}
The lowest term in the expansion around the center of the Dalitz plot
for the decay amplitude squared is proportional to $z$:
\begin{equation}
\mid \bar{A}(z,\phi)\mid^2= c_0(1 + 2\alpha z +...) 
\label{eqn:alpha}
\end{equation}
Experimental  and theoretical  results on  the $\alpha$  parameter are
summarized in table~\ref{tab:table2}. The two recent low background and
high  statistics experiments (about  \EA{6} events  in the  final data
sample):   Crystal   Ball  at   the   AGS  \cite{Tippens:2001fm}   and
KLOE~\cite{Capussela:2005rp}  (2005   preliminary  result),  were  not
consistent until very recently. The discrepancy seems to be removed in
the recent reanalysis of the KLOE data~\cite{Ambrosino:2007wi}.
\begin{table}
\caption{\label{tab:table2} Experimental and theoretical results for 
the parameter $\alpha$.}
\begin{tabular}{l|l|r}
 $\alpha$ & Comment& Ref. \\
\hline
-0.052 $\pm$ 0.017(stat) $\pm$ 0.010(syst)& Exp (CBarrel)  & \cite{Abele:1998yi}     \\
-0.031 $\pm$ 0.004                        & Exp (CBall)  & \cite{Tippens:2001fm}     \\ 
-0.013 $\pm$ 0.004(stat) $\pm$ 0.005(syst)& Exp (KLOE), prel.  & \cite{Giovannella:2005rz}  \\
-0.027 $\pm$ 0.004(stat) $\pm$ $^{+0.005}_{-0.006}$(syst)& Exp (KLOE), prel.  & \cite{Ambrosino:2007wi}  \\
-0.026 $\pm$ 0.01(stat) $\pm$ 0.01(syst)  & Exp (WASA)  & \cite{Bashkanov:2007iy}    \\
\hline
0                                         & Current Algebra  & \cite{PhysRevLett.18.1170}       \\
+0.015                     & CHPT,1loop   & \cite{Gasser:1984pr,Bijnens:2002qy}      \\
-0.007...-0.014                           & CHPT$+$dispersive  & \cite{Kambor:1995yc}\\
-0.007                                    & UCHPT  & \cite{Beisert:2003zs}           \\
-0.031 $\pm$ 0.003                        & UCHPT/fit  & \cite{Borasoy:2005ws}       \\
\end{tabular}
\end{table}

A  consequence of the  fact that  $m_{\pi^0}<m_{\pi^\pm}$ is  that the
kinematical region  of the $\eta\to  3\po$ decay phase  space includes
the threshold  for the $\pio\pio\to\pip\pim$  process.  The boundaries
corresponding  to the  \pio\pio\ invariant  mass  ($M_{\pi\pi}$) equal
$2m_{\pi^+}$  are illustrated by  dashed lines  in Fig.~\ref{fig:def}.
Recently a  cusp in  $M_{\pi\pi}$ was observed  in $K^+\to\pi^+\po\po$
decay  by  the  NA48/2  collaboration  \cite{Batley:2005ax}.   Initial
interpretation    of    the    effect    was    given    by    Cabibbo
\cite{Cabibbo:2004gq}  and   the  process  could   provide  a  precise
determination of a combination of the $a_0$, $a_2$ the $I=0$ and $I=2$
pion-pion  $s$-wave scattering  lengths.  A  similar effect  should be
present    in    the     $K_L\to    3\po$,    $\eta\to    3\po$    and
$\eta'\to\pio\pio\eta$ decays.

The cusp effect in $\eta\to 3\pio$  has been studied in a one loop
approximation by Belina \cite{Belina:2006bb}. The following expression
for the amplitude is obtained:
\begin{equation}
\bar{A}(s_1,s_2,s_3)\propto 1+k_1\sum_{i=1}^3\bar{J}(s_i;m_{\po})+
k_2\sqrt{\frac{2}{3}}\sum_{i=1}^3\bar{J}(s_i;m_{\pim})
\end{equation}
where
\begin{equation}
16\pi^2 \bar{J}(s,m)=\left\{\begin{array}{lcl}
2-2\sqrt{-\Delta}\arctan\sqrt{-\frac{1}{\Delta}}&{;}&0<s<4m^2\\
2+\sqrt{\Delta}\left[\ln\frac{1-\sqrt{\Delta}}
{1+\sqrt{\Delta}}+i\pi\right]&{;}&4m^2<s\\
\end{array}
 \right.
\end{equation}
and  $\Delta=1-\frac{4m^2}{s}$.  The $k_1$  and  $k_2$ parameters  are
related to the $\pi\pi$ scattering lengths $a_0$ and $a_2$:
\begin{equation}
  k_2=\frac{32\pi}{3}(a_0-a_2); \ \   k_1=\frac{16\pi}{3}(a_0+2a_2).
\end{equation}
Here the values $a_0-a_2=0.265$ and  $a_0+2a_2=0.13$ will be used in order
to estimate the magnitude of the effect.

\begin{figure}
     \includegraphics[width=0.5\textwidth,clip]{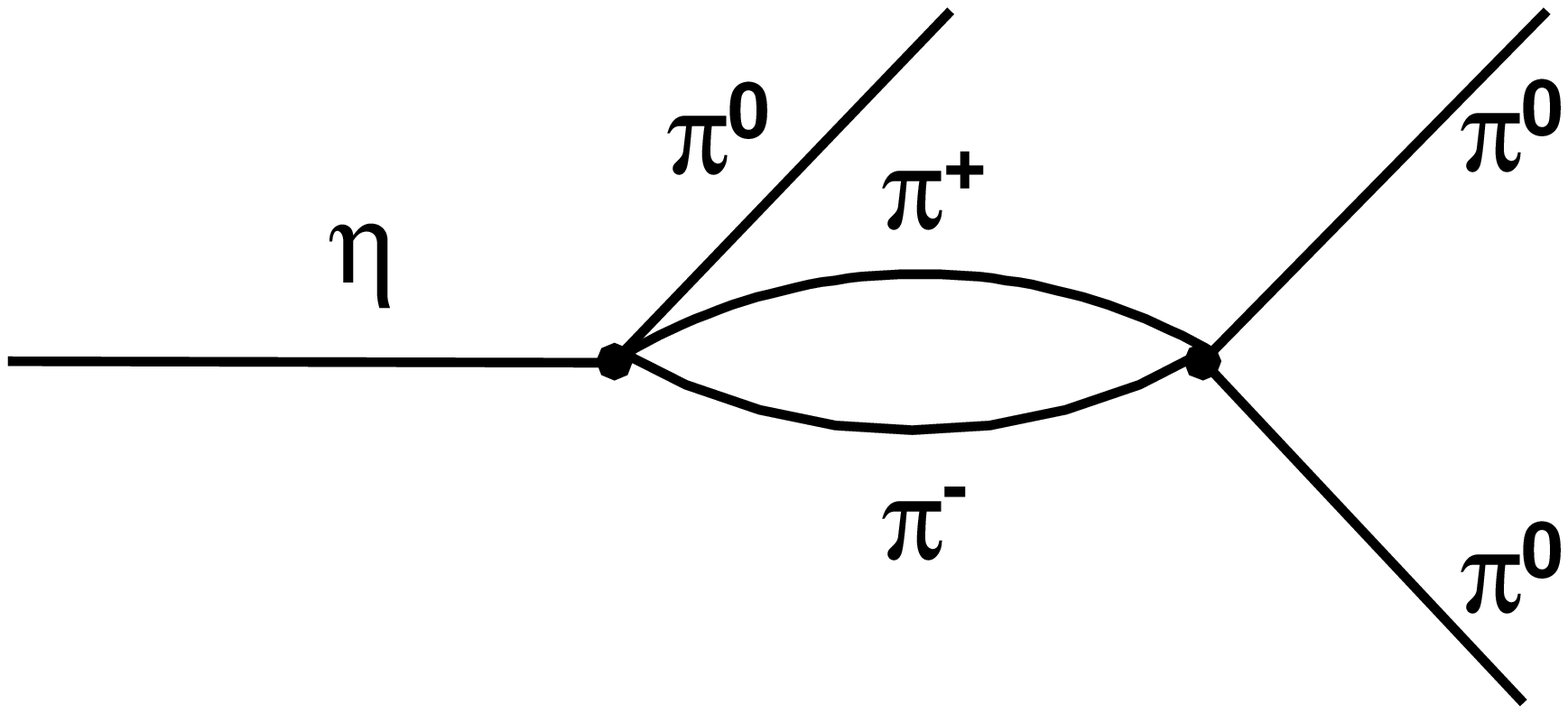}
     \caption{\label{fig:diagcusp}             Lowest            order
$\pi^+\pi^-\to\pi^0\pi^0$    rescattering    contribution    in    the
$\eta\to\pio\pio\pio$ decay.}
\end{figure}

\begin{figure}
     \includegraphics[width=0.49\textwidth,clip]{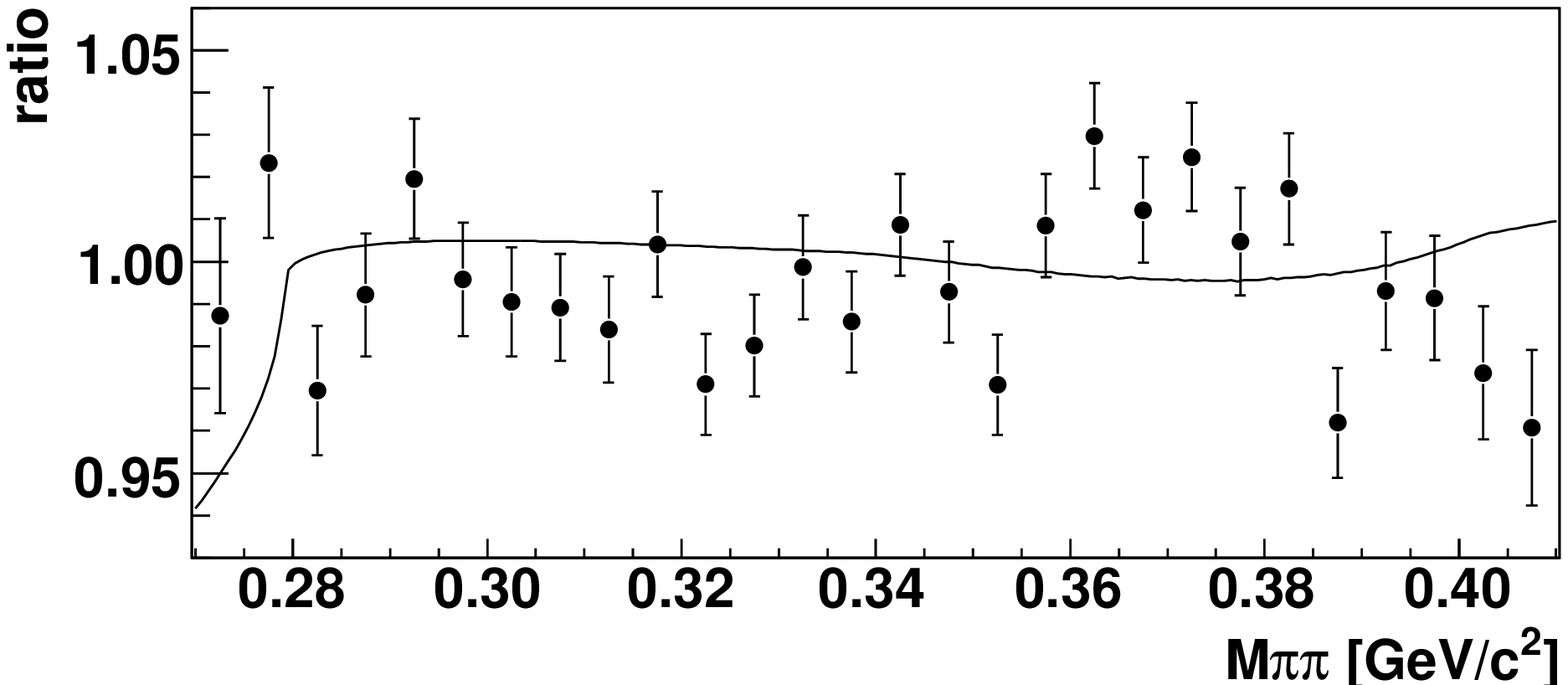}
     \includegraphics[width=0.49\textwidth,clip]{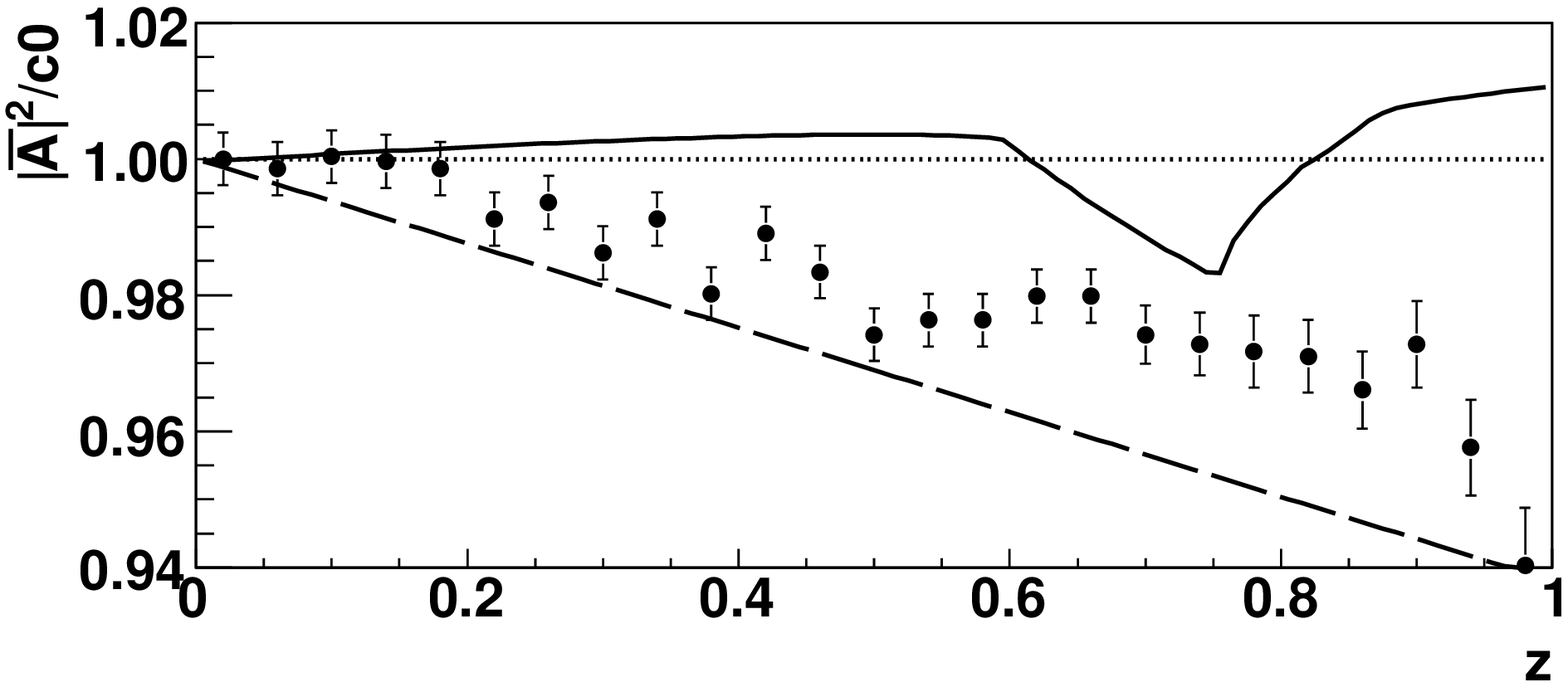}
     \caption{\label{fig:cusp}       (left)      $\mathrm{M}_{\pi\pi}$
distribution (divided  by the phase space): line  -- CHPT calculations
from Belina \cite{Belina:2006bb},  points -- CELSIUS/WASA experimental
results  \cite{Bashkanov:2007iy}. (right)  $|\bar{A}(z)|^2/c_0$: solid
line  --  Belina;  dashed  line  -- $\alpha=-0.031$;  dotted  line  --
$\alpha=0$;     points    --     preliminary     data    from     KLOE
\cite{Ambrosino:2007wi}.}
\end{figure}

The  predictions  for the  $M_{\pi\pi}$  distribution  divided by  the
$\eta\to 3\pio$ phase space in  the region below the $2m_{\pi^+}$ mass
(279  MeV)  reveals  a deviation  up  to  few  percent (left  part  of
Fig.~\ref{fig:cusp}).   The observation  of the  cusp would  require a
resolution of  a few MeV/c$^2$  in $M_{\pi\pi}$ (e.g.   the resolution
obtained by CELSIUS/WASA  of 6 MeV/c$^2$ (RMS) would  be enough) and a
data      sample     of      about     \EA{7}      $\eta\to     3\pio$
decays~\cite{Bashkanov:2007iy}.  Since  the required data  samples are
only a few  times larger than these collected so  far by Crystal Ball,
KLOE  or WASA-at-COSY~\cite{Vlasov:2007aa} one should expect  that the
effect will be observed in the near future.
 
Another aspect of  the cusp effect is how the  $z$ distribution and in
the  result the extracted  $\alpha$ parameter  could be  affected. The
predictions        from       \cite{Belina:2006bb}        are       in
Fig.~\ref{fig:cusp}(right) given by solid  line.  The influence of the
cusp shows up in the  region of $0.6<z<0.9$ with the maximal deviation
of 2\% for $z\approx 0.75$.   One can also observe that $\alpha>0$ for
one loop CHPT as given  in table~\ref{tab:table2}.  For a comparison a
line represents the  Crystal Ball~\cite{Tippens:2001fm} result and the
points      are      extracted      from     the      newest      KLOE
paper~\cite{Ambrosino:2007wi}  (from  the  Fig.~4).  It  is  therefore
clear  that  the  cusp  effect  cannot  be  ignored  for  the  precise
determination of $\alpha$. One can for example exclude the cusp region
by fitting $\alpha$ only for $0\le z\le0.6$ or include the theoretical
predictions to describe the effect.

In   addition   to  the   cusp,   at   the   vicinity  of   the   same
$M_{\pi\pi}=2m_{\pi^+}$ boundary pionium ($\pi^+\pi^-$ atom) should be
created.   Pionium  was  observed  in  the DIRAC  experiment  and  its
lifetime   measured  \cite{Adeva:2005pg}   leading  to   an  $a_0-a_2$
determination.  The production rate of  the pionic atoms in the decays
of   $K$,  $\eta$,   $\eta'$   was  studied   by   Wycech  and   Green
\cite{Wycech:1993ci}  and  Silagadze  \cite{Silagadze:1994wr}.   Since
pionium     decays      predominantly     into     \pio\pio\     pair:
$\Gamma\approx\Gamma_{\pi^0\pi^0}$                 (\cite{Deser:1954vq,
Gasser:2001un})    and    therefore     should    be    observed    in
$\eta\to\pio\pio\pio$,     $\eta'\to\pio\pio\eta$     and    $K^\pm\to
\pio\pio\pi^\pm$ decays:
\begin{equation}
\frac{\Gamma(P_0 \to A_{2\pi}P_3)}{\Gamma(P_0 \to \pi^+ \pi^- P_3)}=
\frac{\pi}{R}\alpha_{\mathrm{em}}^3 \frac{m^2}{m_0^2} \vert 
A(\tilde x,\tilde y) \vert^2
\sqrt{\frac{1}{4}\left(1-4\frac{m^2}{m_0^2}+\frac{m_3^2}
{m_0^2}\right)^2-\frac{m_3^2}{m_0^2}},
\label{eq5}
\end{equation}
where $\tilde x$ and $\tilde y$ are Dalitz variables, corresponding to
the atom creation:
\begin{equation}
\tilde x=0, \hspace*{5mm} \tilde y=\frac{3}{2}
\frac{m_0^2-4m^2+m_3^2}{m_0(m_0-2m-m_3)}
\end{equation}
R is a dimensionless remnant of the three-particle phase space
integral defined in \cite{Silagadze:1994wr}.  The result for the
branching ratios is:
\[
BR(\eta \to  A_{2\pi}\pi^0) \approx 2 \cdot 10^{-8};
BR(\eta^\prime \to  A_{2\pi}\eta) \approx 6 \cdot 10^{-7}; 
BR(K^\pm \to  A_{2\pi}\pi^\pm) \approx 5.5 \cdot 10^{-7}.
\]
That corresponds  to the fractions:
\[
\frac{\Gamma(\eta \to  A_{2\pi}\pi^0)}{\Gamma(\eta \to 2\pi^0\pi^0)} 
\approx 6\cdot 10^{-8};
\frac{\Gamma(\eta^\prime \to A_{2\pi} \eta)}{\Gamma(\eta' \to 2\pi^0\eta)} 
\approx 3 \cdot 10^{-6};
\frac{\Gamma(K^\pm \to  A_{2\pi}\pi^\pm)}{\Gamma(K^\pm\to 2\pi^0\pi^\pm)} 
\approx 3 \cdot 10^{-5}.
\]
In  case of  $K^\pm\to\pi^0\pi^0\pi^\pm$ statistics  $\sim$  \EA{8} is
available  and there  are  some  hints that  the  pionium is  observed
\cite{Goudzovski:2007wx}.   The  signature would  be  a  tiny peak  in
$M_{\pi\pi}$ distribution at $M_{\pi\pi}=2m_{\pi^\pm}$.

\section{Sources  of the  $\eta$ and $\eta'$ mesons}

The  \et\ and  \etp\ mesons  can  be copiously  produced in  $\pi^-p$,
$\gamma  p$, $pp$  or $pd$  interactions not  too far  from  the meson
production thresholds.  The mesons  in this case originate from decays
of nucleon  isobars (\et\ mainly  from $N^*(1535)$).  Crystal  Ball at
BNL    (Brookhaven)    employed    $\pi^-   p\to    n\eta$    reaction
\cite{Prakhov:2005qb}  and after  the  move of  the  detector to  MAMI
(Mainz) \cite{Starostin:2005rq} the \et\ and \etp\ mesons are produced
in the $\gamma p\to p\eta,\eta'$ reaction close to threshold.  In WASA
at CELSIUS the \et\  mesons were produced in $pd\to ^3\mathrm{He}\eta$
at    threshold~\cite{Bargholtz:2006gz}   and   in    $pp\to   pp\eta$
reactions~\cite{Calen:1996mn}.    In  WASA-at-COSY  $\eta'$   will  be
produced    in     $pp\to    pp\eta'$    reaction~\cite{Moskal:1998pc,
Moskal:2000gj}.

Radiative decays  of $\phi$ mesons  produced at the resonance  peak in
$e^+e^-$  colliders are  used by  CMD-2~\cite{Akhmetshin:1997tz}, SND,
KLOE~\cite{Aloisio:2002ac}   experiments.   In   the   Crystal  Barrel
experiment  the mesons  were produced  in $p\bar{p}$  annihilations at
rest.  In  addition there are  high energy hadronic reactions  such as
peripheral  $\pi^-p\to n\eta,\eta'$~\cite{Bolotov:1974vy, Apel:1979ic}
production used  in experiments  on meson spectroscopy  performed with
GAMS2000  spectrometer~\cite{Binon:1985ag}  and  recently by  the  VES
collaboration~\cite{Dorofeev:2006fb}.  A unique possibility to measure
the radiative width of the  mesons is provided by \et, \etp\ formation
in  $\gamma(^*)\gamma(^*)$  processes  as  e.g.  in  the  $e^-e^\pm\to
e^-e^\pm\eta,\eta'$ reaction.

\begin{table}
\caption{\label{tagging}  Parameters of  close to  threshold  \et\ and
\etp\ meson production reactions. $T_b$, $p_b$ are beam kinetic energy
and momentum and $\beta$ velocity of  the CMS at threshold. $Q$ is CMS
excess energy corresponding to an  {\em optimal} beam energy and $\sigma$
is  reaction cross  section.   The last  column $\sigma_T$  indicates
total inclusive  cross section of  the processes originating  from the
given initial state. }
\begin{tabular*}{0.9\textwidth}{@{\extracolsep{\fill}}l|rrrrrr}
\hline
&$T_b$&$p_b$&$\beta$&Q&$\sigma$ ($\sigma_{max}$)&$\sigma_T$\\
\hline
$pp\to pp\eta$      & 1.253& 1.981&0.63&40&10$\mu$b&40mb    \\
$pp\to pp\eta'$     & 2.404& 3.208&0.75 &45&300nb&40mb\\
$pd\to ^3$He$\eta$  & 0.891& 1.569&0.42&2&400nb&80mb\\
$\pi^- p\to n\eta$  & 0.559& 0.684&0.42&&2.8 mb&50mb\\
$\gamma p\to p\eta$ & 0.706& 0.706&0.43&58&16$\mu$b&300$\mu$b\\
$\gamma p\to p\eta'$& 1.447& 1.447&0.61&27$^*$&1$\mu$b&150$\mu$b\\
\hline
\end{tabular*}
\end{table}

A feature of the close  to threshold processes is that the particle(s)
that accompanies  the produced  meson are emitted  in a  small forward
cone  and their  momenta can  be  measured accurately  in a  dedicated
detector which  covers a limited  range of scattering angles.   A good
resolution in the missing mass  can therefore be achieved (typically a
few MeV/c$^2$  FWHM) and  provide a clear  identification of  \et\ and
\etp\ production.  The decay  particles are emitted more isotropically
(since the velocity of the center of mass system is not very high) and
their registration requires a detector with nearly 4$\pi$ sr coverage.
A typical resolution for the invariant masses of the decay products is
a few ten  MeV/c$^2$. In the close to  threshold processes the tagging
particles and the decay particles  appear in separate regions of phase
space  and  branching  ratios  $\Gamma_i/\Gamma_\mathrm{tot}$  can  be
determined directly.

Among the close  to threshold reactions \et, \etp\  production in $pp$
and $pd$ processes  have the lowest ratio of the  cross section to the
total  inclusive.   The signal  to  background  ratio  can however  be
enhanced  by increasing  the  resolution  in the  missing  mass or  by
decreasing the  beam energy  to work closer  to the threshold  (at the
price of  a lower production  cross section). The ideal  reaction from
this respect is $pd\to^3$He$\eta$, where the cross raises very quickly
reaching 0.4 $\mu$b already at excess  energy of about 1 MeV above the
threshold~\cite{Mayer:1995nu,     Mersmann:2007gw,    Smyrski:2007nu}.
Together with  the fact  that $\eta$ is  accompanied by one  heavy and
doubly  charged particle  this allows  for  a very  clean and  precise
tagging   by  identification  of   the  $^3$He   ion  in   a  magnetic
spectrometer.   To  cover  the   full  phase  space  the  spectrometer
acceptance below  one degree is sufficient.   The $pp\to pp\eta,\eta'$
reactions can provide the highest useful rate of tagged \et\ and \etp\
mesons  enabling studies  of the  rare  decays which  have a  distinct
signature  like  for  example  the  decay  $\et\to\elp\elm$  where  an
integrated luminosity corresponding to \EA{10} \et s is needed.

For  \et\  and  \etp\  originating  from $\phi$  radiative  decays  in
$e^+e^-$ colliders the general level  of background is much lower than
in the hadronic interactions.  The physics background is mainly due to
the other $\phi$  decays: $BR(\phi\to\eta\gamma)=1.301\pm 0.024\%$ and
$BR(\phi\to\eta'\gamma)=6.2\pm  0.7\E{-5}$.   The pseudoscalar  mesons
are produced  antiparallel to a monoenergetic photon  (with energy 360
MeV for  \et\ and 60  MeV for \etp)  and have relatively low  and well
defined velocity.  The monoenergetic  photon gives in principle a very
clean  signature.   However for  neutral  decays  this  photon can  be
misidentified  with photons  from  \et(\etp) since  the resolution  of
present electromagnetic calorimeters does not provide stringent enough
constraints  for the \et(\etp)  fourmomentum.  When  charged particles
are produced,  the decay system can  be measured much  better. At KLOE
for  example the  invariant mass  resolution for  a system  of charged
particles is  comparable to  the $pp$ missing  mass resolution  in the
WASA detector.

The  $e^+e^-$  colliders  also  allow  production  of  the  mesons  in
$\gamma\gamma$ interactions by  the $e^+e^-\to e^+e^- \gamma^*\gamma^*
\to e^+e^-\eta,\eta'$ process.  The  cross section depends on the flux
of  the  virtual  photons   and  cross  section  of  the  $\eta,\eta'$
production  which  is  proportional  to  the radiative  width  of  the
mesons~\cite{Budnev:1974de}.   In all  experiments so  far  the mesons
were  identified by  detecting a  specific decay  mode that  allows to
measure the $\Gamma(\gm\gm)\times \Gamma(i)/\Gamma_{tot}$ combination.
All results on  \et\ and \etp\ radiative widths  were obtained by this
indirect method~\cite{Yao:2006px}.  The  mesons are produced by almost
real virtual  photons and the  outgoing electrons are emitted  at very
small angles. The energy of  the electron and the positron will differ
from the  beam energy  and therefore they  will be deflected  from the
beam region  by the collider dipoles.  The  deflection angles measured
in specialized  position detectors will determine  their energies.  If
both electron and positron are  measured one can tag the production of
$\eta,\eta'$ without having to rely on a decay channel.  For the KLOE2
experimental programme  it is planned  to equip the detector  with two
tagging   spectrometers   and   the   first  direct   measurement   of
$\Gamma(\et,\et'\to\gm\gm)$  would be possible.   In order  to produce
\etp\ mesons and to avoid background due to the $\phi$ decays a higher
center of mass energy is required.

In conclusion one should stress  that studies of \et\ and \etp\ decays
require high statistics precision experiments. A good understanding of
systematical  effects  can  be  obtained  by  comparing  results  from
experiments that use different techniques.

\begin{theacknowledgments}
Many thanks to Pawe{\l} Moskal and his crew for the invitation to this
excellently organized symposium.
\end{theacknowledgments}

\bibliographystyle{aipproc}   

\bibliography{oldbib,etaov}
\end{document}